\documentclass[conference]{IEEEtran}
\IEEEoverridecommandlockouts

\usepackage[utf8]{inputenc}
\usepackage[T1]{fontenc}
\usepackage{comment}
\usepackage{cite}
\usepackage{amsmath,amssymb,amsfonts}
\usepackage{algorithmic}
\usepackage[linesnumbered,ruled]{algorithm2e}
\usepackage{graphicx}
\usepackage{textcomp}
\usepackage{xcolor}
\usepackage{booktabs}
\usepackage{listings}
\usepackage{mdframed}
\usepackage{cleveref}
\usepackage{breqn}
\usepackage{stfloats}
\usepackage{afterpage}
\usepackage{float}
\usepackage{here}
\usepackage{caption}
\usepackage{subcaption}
\usepackage{url} 

\usepackage[most]{tcolorbox} 
\def\BibTeX{{\rm B\kern-.05em{\sc i\kern-.025em b}\kern-.08em
    T\kern-.1667em\lower.7ex\hbox{E}\kern-.125emX}}

\usepackage{tikz}
\usepackage{lipsum} 
\makeatletter
\def\ieeecopyright{
  \footnotesize
  © 2025 IEEE. Personal use of this material is permitted.}
\makeatother
\AddToHook{shipout/firstpage}{%
  \begin{tikzpicture}[remember picture,overlay]
    \node[anchor=south west,xshift=1.0cm,yshift=0.8cm]
      at (current page.south west)%
      {\parbox{\linewidth}{\raggedright\ieeecopyright}};
  \end{tikzpicture}%
}

\crefname{figure}{Fig.}{Figures}
\crefname{table}{TABLE}{Tables}
\crefname{section}{Section}{Sections}
\crefname{equation}{eq.}{eqs.}

\begin{document}

\newcommand{\TOOLNAME}{\textit{Agnocast}}

\title{
ROS~2 Agnocast: Supporting Unsized Message Types for True Zero-Copy Publish/Subscribe IPC
}

\author{
\IEEEauthorblockN{Takahiro Ishikawa-Aso and Shinpei Kato}
\IEEEauthorblockA{The University of Tokyo}
\IEEEauthorblockA{TIER IV, Inc.}
}

\lstdefinestyle{mystyle}{
    basicstyle=\ttfamily\footnotesize,
    breakatwhitespace=false,
    breaklines=true,
    captionpos=b,
    keepspaces=true,
    numbers=left,
    numbersep=10pt, 
    showspaces=false,
    showstringspaces=false,
    showtabs=false,
    tabsize=2,
    framexleftmargin=15pt, 
    xleftmargin=15pt,
    morekeywords={agnocast, message_ptr}, 
    keywordstyle=\color{red}\bfseries, 
}

\lstset{style=mystyle}

\newtcolorbox{codebox}[1][]{
    enhanced,
    unbreakable,
    colback=white,
    colframe=black,
    boxrule=0.5pt,
    sharp corners,
    boxsep=2pt,
    left=5pt,
    right=5pt,
    top=2pt,
    bottom=2pt,
    listing only,
    listing options={style=mystyle},
    overlay={\node[anchor=south east, outer sep=2pt, font=\footnotesize] at ([xshift=-3pt, yshift=3pt]frame.south east) {#1};}
}

\maketitle

\begin{abstract}
Robot applications, comprising independent components that mutually publish/subscribe messages, are built on inter-process communication (IPC) middleware such as Robot Operating System~2 (ROS~2).
In large-scale ROS~2 systems like autonomous driving platforms, true zero-copy communication—eliminating serialization and deserialization—is crucial for efficiency and real-time performance.
However, existing true zero-copy middleware solutions lack widespread adoption as they fail to meet three essential requirements:
1) Support for all ROS~2 message types including unsized ones;
2) Minimal modifications to existing application code;
3) Selective implementation of zero-copy communication between specific nodes while maintaining conventional communication mechanisms for other inter-node communications including inter-host node communications.
This first requirement is critical, as production-grade ROS~2 projects like Autoware rely heavily on unsized message types throughout their codebase to handle diverse use cases (e.g., various sensors), and depend on the broader ROS~2 ecosystem, where unsized message types are pervasive in libraries.
The remaining requirements facilitate seamless integration with existing projects.
While IceOryx middleware, a practical true zero-copy solution, meets all but the first requirement, other studies achieving the first requirement fail to satisfy the remaining criteria.
This paper presents \TOOLNAME{}, a true zero-copy IPC framework applicable to ROS~2 C++ on Linux that fulfills all these requirements.
Our evaluation demonstrates that \TOOLNAME{} maintains constant IPC overhead regardless of message size, even for unsized message types.
In Autoware PointCloud Preprocessing, \TOOLNAME{} achieves a 16\% improvement in average response time and a 25\% improvement in worst-case response time.
\end{abstract}

\vspace{1mm}


\vspace{-2mm}

\section{Introduction}
With recent advancements in computing technology, large-scale robotic applications, including autonomous driving systems, have become vital infrastructures in modern society.
Robot applications are composed of independent components that interact through a message-passing mechanism on publish/subscribe communication middleware, such as Robot Operating System 2 (ROS~2) \cite{ros2doc}.
ROS~2 is a de-facto standard platform underlying many robot applications today.
The performance of Inter-Process Communication (IPC), foundational to the publish/subscribe mechanism, is critical for minimizing latency and ensuring overall system efficiency.
High IPC latency can deteriorate the response times of robotic applications, and the increased CPU and memory costs associated with IPC negatively impact overall system throughput and exacerbate memory pressure.
To minimize IPC cost, eliminating copies including serialization and deserialization is ideal, which we call \textbf{true zero-copy}.
In ROS~2, true zero-copy communication can be achieved by placing nodes within the same process through the \textit{ComponentContainer} feature~\cite{macenski2023impact}.
However, this method compromises fault isolation.
For example, in autonomous driving systems, numerous nodes work collaboratively to achieve functionalities such as localization and object recognition.
When fault isolation is not implemented through process separation between nodes, the failure of a single node's process can lead to complete system failure.
Therefore, there is a strong motivation to implement true zero-copy communication that enables node distribution across separate processes without sacrificing IPC performance.

Although practical true zero-copy middleware already exists, it is restricted to handling messages of static size.
IceOryx \cite{pohnl2022middleware, pohnl2023shared} is a practical true zero-copy middleware that allows direct access from the ROS~2 standard interface when using CycloneDDS \cite{cyclonedds}.
However, its limitation to handling only ROS~2 messages of static size necessitates the rewriting of message definition files for each project to hardcode static array sizes.
Additionally, any existing source code that assumes unsized arrays (e.g. \lstinline|std::vector| in C++) must also be modified to accommodate the static array requirements.
Many libraries in the ROS 2 community are designed under the assumption that message types are unsized.
Autoware \cite{kato2015open, kato2018autoware}, one of the large-scale ROS~2 applications, relies heavily on unsized message types throughout their codebase to handle diverse use cases (e.g., various sensors) and to utilize broader ROS~2 ecosystem, where unsized message types are pervasive.

\begin{figure}[tb]
  \centering
  \includegraphics[width=1.00\linewidth]{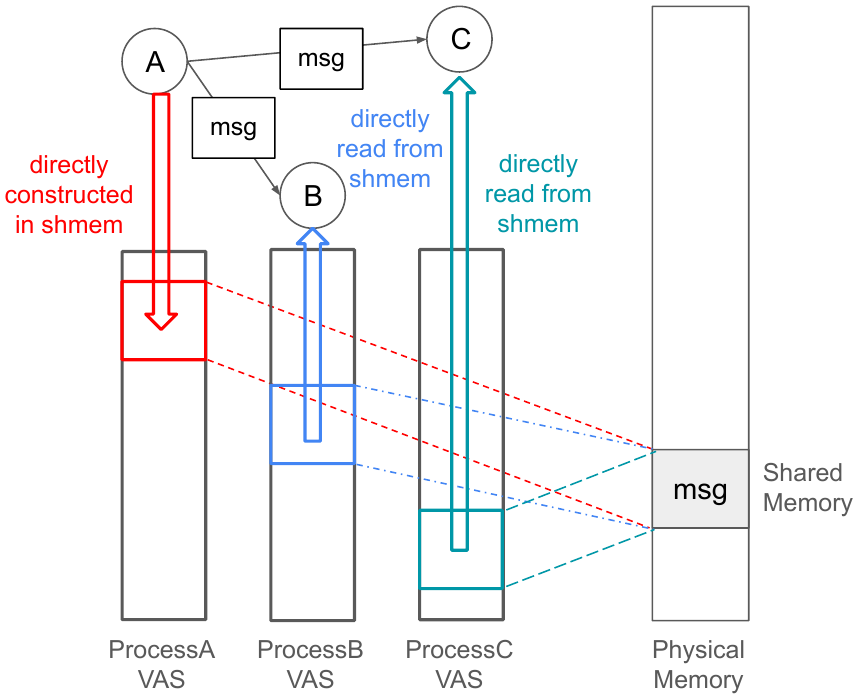}
  \caption{True zero-copy IPC in publish/subscribe communication. Process A constructs messages directly in shared memory, while Processes B and C read them directly from the shared memory without serialization or deserialization.}
  \label{fig-true-zero-copy}
  \vspace{-5mm}
\end{figure}

True zero-copy publish/subscribe IPC for unsized message types has involved trade-offs with existing middleware standards compliance.
This means it has required substantial rewriting of existing user code and surrounding libraries.
Research has explored middleware for ROS~1 and ROS~2 that enables true zero-copy IPC for unsized message types, such as TZC \cite{wang2019tzc} and LOT \cite{iordache2021smart}.
However, these approaches fall short of practical application in real-world projects due to limitations in standards compliance and usability.
As depicted in \cref{fig-true-zero-copy}, achieving true zero-copy communication while maintaining fault isolation typically necessitates directly constructing all message data in shared memory.
This, in turn, often requires library users to be aware of, and directly manage, shared memory allocation.
TZC and LOT, while achieving true zero-copy for unsized message types, impose significant constraints on application code.
TZC's partial serialization approach requires custom message type implementations and pre-allocation of array sizes.
LOT, similarly, requires conversion to custom message formats and imposes library-specific coding constraints.
These limitations not only increase development overhead but also severely restrict the flexibility of application development by preventing the free use of methods like \lstinline|std::vector::push_back|, which may require reallocation at arbitrary times.
Existing publish/subscribe middleware, including ROS~2 and its underlying rmw implementations~\cite{cyclonedds, fastdds, connext, zenoh}, was not designed with true zero-copy support for unsized message types in mind, further compounding the challenge.

This study introduces a true zero-copy framework, named \textbf{\TOOLNAME{}}, applicable to ROS~2 C++ on Linux platforms.
It supports all ROS~2 message types including unsized message types, enabling true zero-copy communication without compromising ROS~2 compliance.
Adapting \TOOLNAME{} requires changes only to the ROS~2 publish/subscribe APIs and a smart pointer type for the IPC messages.
We first identify a constraint within the C++ language ecosystem and Interface Definition Language (IDL) that makes it impossible to specify a memory allocation strategy for ROS~2 message objects.
This constraint impedes the allocation of message objects with unsized types in shared memory while adhering to ROS~2 standards.
To address this issue, \TOOLNAME{} intercepts all heap allocations using the LD\_PRELOAD functionality and makes the entire heap area accessible to subscriber processes in a read-only mode.
In addition to enabling zero-copy communication, our \textit{bridge} feature allows ROS~2 communication and \TOOLNAME{} communication to be mutually relayed, enabling specific edges (communication between two specific nodes) to utilize true zero-copy while allowing the entire ROS~2 system to maintain its original functionality.
\TOOLNAME{} is independent of the specifics of existing middleware implementations in ROS~2, such as CycloneDDS \cite{cyclonedds}, FastDDS \cite{fastdds}, RTI Connext \cite{connext}, and Zenoh \cite{zenoh}, allowing it to remain agnostic to changes in \textit{rmw} implementation \cite{rmw}.
The \TOOLNAME{} IPC framework's concept is not tied to ROS~2 and can be adapted for other middleware systems with similar requirements and constraints.

\textbf{Contributions.} The main contributions of this paper include the following:

\begin{enumerate}
\item We propose a novel IPC technology that achieves true zero-copy communication with minimal code modifications through a new smart pointer design, particularly effective for systems unable to specify memory allocation strategies in application code.
\item The \TOOLNAME{} IPC framework supports all ROS~2 message types, covering unsized message types, for true zero-copy publish/subscribe communication without compromising compliance with ROS~2 standards.
\item We develop \TOOLNAME{} for ROS~2 as open-source and demonstrate the performance of the zero-copy feature.
\end{enumerate}

\section{MOTIVATION} \label{section-motivation}
Autoware \cite{kato2015open, kato2018autoware} is the world's leading open-source project for autonomous driving, built on ROS~2 C++, representing a widely adopted autonomous system in real-world applications.
ROS~2's component-based architecture enables node-based development of various functionalities.
These nodes can be flexibly combined to construct autonomous driving systems that meet specific operational requirements.
This paper proposes a design for true zero-copy middleware that is highly applicable to real-world ROS~2 applications.
We analyze Autoware as a representative case study to establish the fundamental background for determining middleware requirements.

Autoware's autonomous driving systems call for true zero-copy IPC for publish/subscribe messaging.
The system combines hundreds of nodes exchanging messages such as large pointcloud data.
While these systems often span multiple hosts, most message communication occurs between processes on the same host.
Message transfer across processes incurs copy costs from serialization/deserialization and underlying communication middleware such as Data Distribution Service (DDS), resulting in increased inter-node latency, CPU usage, and memory pressure.
Therefore, zero-copy communication for same-host processes is essential.
While \textit{ComponentContainer} enables true zero-copy communication through pointer passing within a single process, this approach compromises safety requirements of fault isolation.
Autonomous driving systems require fault isolation between nodes, necessitating node deployment in separate processes.
These factors establish the need for true zero-copy inter-process communication.

Despite the existence of true zero-copy middleware solutions such as IceOryx, their adoption in Autoware has been limited.
This is primarily because Autoware extensively uses unsized message types (e.g., message types containing C++'s \lstinline|std::vector|).
Almost all of the ROS~2 nodes implemented in Autoware \cite{autoware_github} depend on unsized message types.
Existing practical true zero-copy middleware solutions, including IceOryx, only support static-sized message types.
Autoware's prevalent use of unsized message types stems from its design goal of handling multiple use cases within a single codebase.
For example, a sensor data preprocessing node can handle various sensor types with a single implementation.
Furthermore, numerous library assets in the ROS~2 ecosystem depend on standardized unsized message types.
Therefore, facilitating adoption in projects such as Autoware, which heavily utilize unsized ROS~2 message types, requires a true zero-copy middleware that supports unsized message types.

\section{REQUIREMENTS AND CONSTRAINTS} \label{section-requirements-constraints}
\vspace{-1mm}
Based on the requirements listed in \cref{subsection-requirements} and the ecosystem constraints outlined in \cref{subsection-constraints}, we introduce our middleware architecture in \cref{subsection-implementation}.

\vspace{-1mm}

\subsection{Middleware Design Requirements} \label{subsection-requirements}
We design a true zero-copy IPC framework named \TOOLNAME{} with the following requirements.

\begin{enumerate}
\item \textbf{True zero-copy with unsized message types.} \label{req-true-zero-copy}
We do not incur copying costs, including serialization and deserialization when transferring ROS~2 messages between processes.
It supports true zero-copy communication for all ROS~2 message types, including unsized ones.

\item \textbf{ROS~2 Compliant.} \label{req-transparency}
True zero-copy communication can be integrated into ROS~2 applications with minimal modifications to the existing code.
We do not need to rebuild tools or libraries, including generated codes related to ROS~2 messages.

\item \textbf{Continuous operation of the ROS~2 entire system.} \label{req-friendly-rmw}
True zero-copy is applied only at the edges between explicitly targeted pairs of nodes.
The remainder of the topic communication continues as before utilizing existing \textit{rmw} communication stacks.
\end{enumerate}

The first requirement serves as the main purpose of our middleware, while the remaining ones facilitate seamless integration with existing projects.
As explained in \cref{section-motivation}, supporting unsized message types (e.g., message types including \lstinline|std::vector| members) is important for widespread adoption.
The definition of an \textbf{unsized message type} is a type for which memory reallocation can occur at arbitrary times, whereas a \textbf{static-sized message type} is one whose memory size is determined at compile time.
In TZC \cite{wang2019tzc} and LOT \cite{iordache2021smart}, although array sizes in messages can be specified dynamically at runtime, this is permitted only once during initialization.
As they do not support memory reallocations at arbitrary times (e.g., \lstinline|std::vector::resize|), they are not considered to support unsized message types.
The second requirement is to avoid adoption barrier seen in solutions such as TZC and LOT, which require extensive modifications to existing application code and message structures; instead, as illustrated in \cref{fig-api}, our middleware implements zero-copy functionality through modifications to publish/subscribe APIs and smart pointer types only.
Given that some ROS~2 applications involve numerous nodes and topics, the third requirement becomes critical to enable selective and incremental adoption of zero-copy communication, starting with the most bandwidth-intensive edges; furthermore, this requirement ensures seamless integration with existing \textit{rmw} middleware for cross-host communications, as zero-copy can only be implemented between nodes on the same host.

\begin{figure}[tb]
\begin{codebox}[Publisher]
\begin{lstlisting}[]
using T = sensor_msgs::msg::PointCloud2;

agnocast::Publisher<T>::SharedPtr pub;
pub = agnocast::create_publisher<T>("mytopic");

void callback(const T::ConstPtr input) {
  agnocast::message_ptr<T> output = publisher->borrow_loaded_message();
  
  fill(input, output);
  
  pub->publish(std::move(output));
}


\end{lstlisting}
\end{codebox}
\vspace{-13pt} 
\begin{codebox}[Subscriber]
\begin{lstlisting}[]
using T = sensor_msgs::msg::PointCloud2;

void callback(const                          agnocast::message_ptr<T> input) {
    ...
}

agnocast::Subscription<T>::SharedPtr sub;
sub = agnocast::create_subscription<T>("mytopic", ...);
\end{lstlisting}
\end{codebox}
\caption{All needed changes to apply \TOOLNAME{} in ROS~2}
\label{fig-api}
\vspace{-5mm}
\end{figure}

\subsection{Middleware Design Constraints in ROS~2} \label{subsection-constraints}
In designing true zero-copy middleware for ROS~2 C++, a key constraint is the inability to specify custom memory allocators for ROS~2 message objects.

Custom memory allocation for ROS~2 message structures is essential for true zero-copy implementation, as messages must be directly constructed on shared memory.
For static-sized message types, middleware such as IceOryx \cite{pohnl2022middleware, pohnl2023shared} can predetermine the required memory size to construct a message.
However, for unsized messages (e.g. containing \lstinline|std::vector|), additional memory allocations occur at arbitrary times in runtime, such as during \lstinline|std::vector::push_back| operations.
Therefore, the memory allocation strategy must be customized to ensure all message data resides on shared memory.

\begin{figure}[tb]
  \centering
  \includegraphics[width=1.00\linewidth]{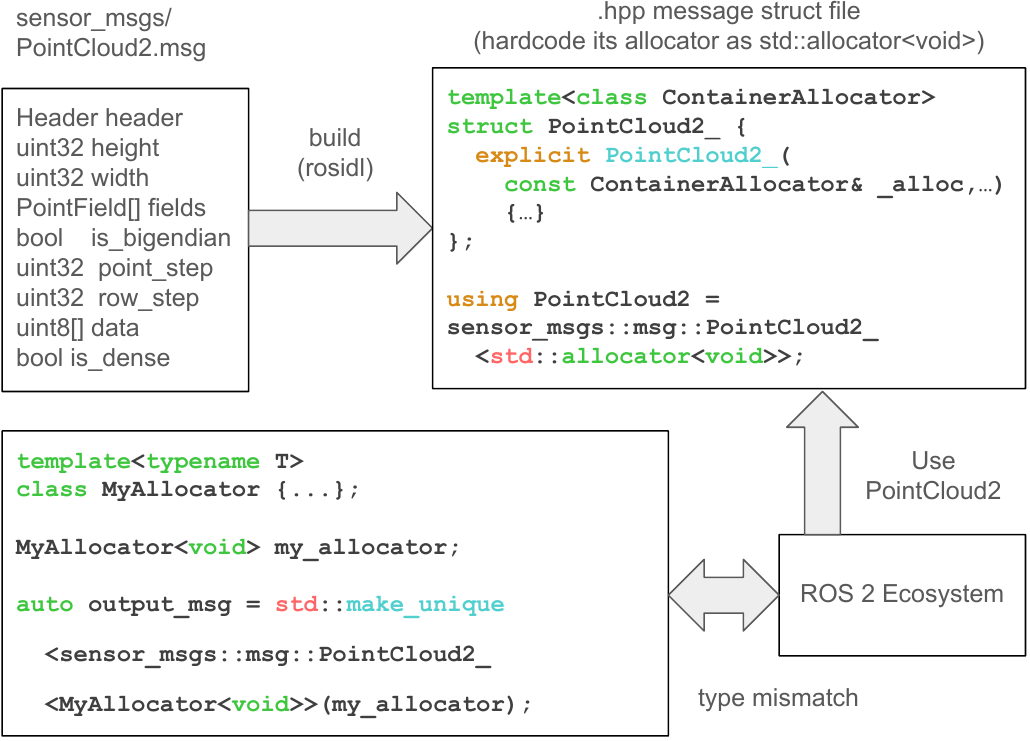}
  \caption{Constraints in ROS~2: hardcoded allocator types in \textit{rosidl} prevent custom memory allocation strategies}
  \label{fig-ros2-ecosystem-constraint}
  \vspace{-3mm}
\end{figure}

However, as depicted in \cref{fig-ros2-ecosystem-constraint}, specifying custom memory allocators for ROS~2 message structures faces significant challenges.
Container types such as \lstinline|std::vector| receive an allocator type as a template argument, seemingly enabling the customization of memory allocation strategies by specifying an allocator that allocates memory from shared memory.
However, the \textit{rosidl} system, responsible for automatically generating source code for ROS~2 message structures, does not allow for the passing of custom memory allocators to container objects within these structures, as the allocator type is hardcoded as \lstinline|std::allocator<void>| in its \textit{empy} template files.
When attempting to implement a custom memory allocator for message structures in the application code, link errors occur due to type incompatibilities between the application code and ROS~2 ecosystem libraries, which expect message types to use the default allocator.

Even C++17's polymorphic allocators (e.g., \lstinline|std::vector<int, std::pmr::polymorphic_allocator<int>>|), which enable allocator specification without changing container types, fail to resolve this issue.
While hardcoding the allocator type to a polymorphic allocator in the \textit{rosidl} code generation process might seem promising, container types using polymorphic allocators are incompatible with existing code that assumes standard container types.
Converting all container types in ROS~2 libraries to use polymorphic allocators could theoretically solve this incompatibility, but such extensive modifications across the entire ecosystem are impractical.

\begin{figure}[tbh]
  \centering
  \includegraphics[width=1.00\linewidth]{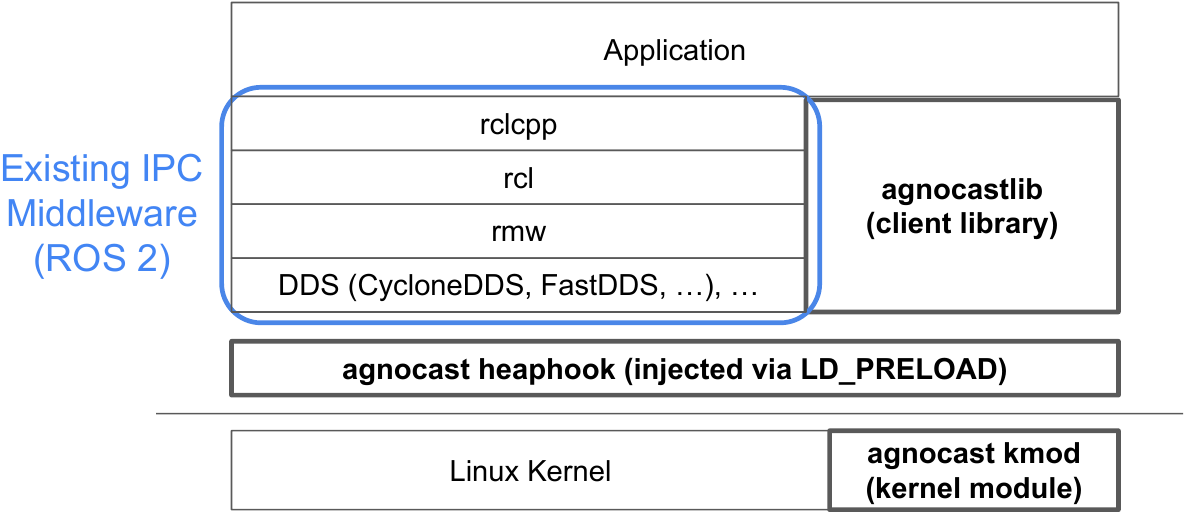}
  \caption{\TOOLNAME{} software stack}
  \label{fig-software-stack}
\end{figure}

\section{AGNOCAST IPC FRAMEWORK} \label{subsection-implementation}
In this section, we detail the design of an IPC framework that addresses the requirements specified in \cref{subsection-requirements}, considering the constraints outlined in \cref{subsection-constraints}. 

\cref{fig-software-stack} illustrates the software stack of \TOOLNAME{} and its integration with existing IPC middleware, such as ROS 2.
\TOOLNAME{} is positioned independently of existing IPC middleware while operating in coordination with it.
As an IPC technology, \TOOLNAME{} can be applied to any component-oriented real-time system with similar requirements and constraints in \cref{section-requirements-constraints}.
Specifically, \TOOLNAME{} serves as a versatile technology for enabling true zero-copy IPC in scenarios where modifying the memory allocation strategy for IPC objects is constrained.

\begin{figure}[h]
  \centering
  \includegraphics[width=1.00\linewidth]{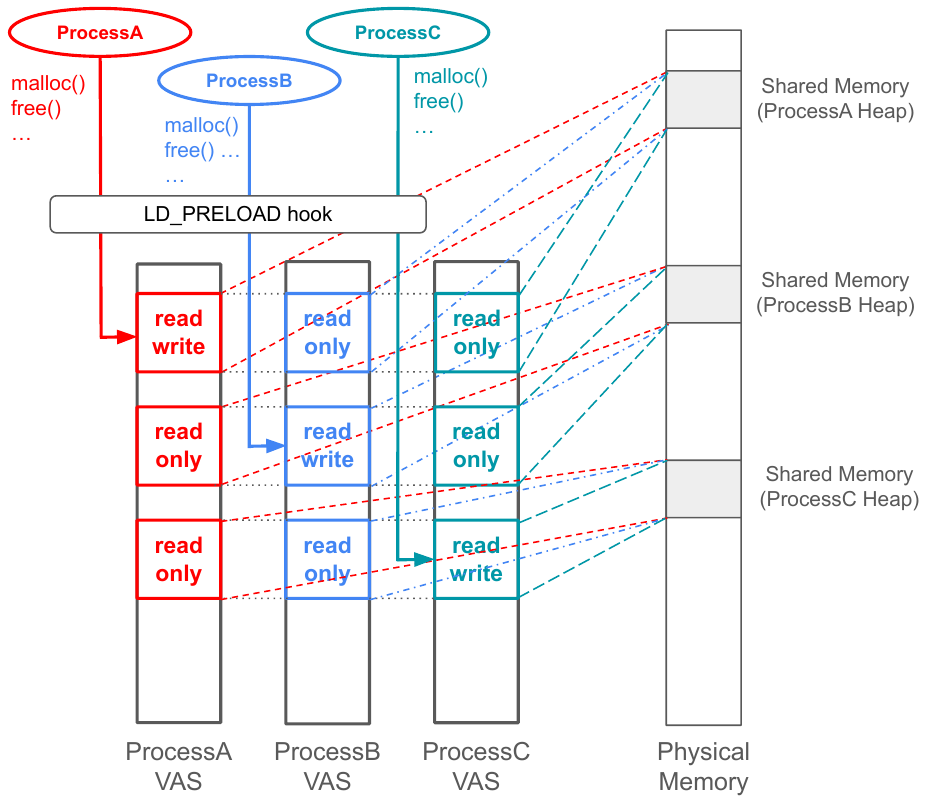}
  \caption{\TOOLNAME{} memory mapping}
  \label{fig-memmap}
  \vspace{-2mm}
\end{figure}

\subsection{\TOOLNAME{} Memory Mapping} \label{subsection-memmap}
Given the constraint that container types like \lstinline|std::vector| cannot receive memory allocation strategies, maintaining ROS~2 compliance while placing message objects in shared memory necessitates replacing the entire heap allocation system.
Virtual memory allocation in C++ container types is controlled by two layers: the allocation strategy passed as template arguments at the container level, and the underlying heap allocation runtime corresponding to \textit{malloc} and \textit{free} functions.
Since the top-level allocation strategy cannot be utilized in ROS~2 as discussed in \cref{subsection-constraints}, we opt to customize the lower layer.
Customizing the heap allocation runtime involves replacing \textit{malloc} and \textit{free} functions to handle shared memory regions instead of standard heap memory.

\cref{fig-memmap} illustrates memory mapping in \TOOLNAME{} using three participating processes as an example.
The diagram shows how the Memory Management Unit (MMU) maps the Virtual Address Spaces (VAS) of three processes to shared memory in physical memory.
The memory mapping in \TOOLNAME{} processes has three key characteristics:

\begin{enumerate}
    \item All heap allocations and deallocations are hooked via LD\_PRELOAD and redirected to operations within a designated virtual address range.
    \item This designated virtual address range is mapped to shared memory in physical memory.
    \item Subscriber processes map the publisher process’s heap into their VAS at the same offset in read-only mode, prohibiting writes to the publisher’s heap.
\end{enumerate}

This memory mapping enables subscriber processes to directly reference any objects placed in the publisher process's heap.
In this memory mapping, true zero-copy IPC is achieved by simply passing object addresses from publisher to subscriber processes, as pointers remain valid when referenced objects maintain identical virtual addresses across processes.
We assume a fixed heap size on shared memory; however, since this refers to a virtual address range rather than physical memory, it can be set sufficiently large.
Since \TOOLNAME{} explicitly specifies the starting address for mmap, and no pointers to outside shared memory are created within the shared memory, Address Space Layout Randomization (ASLR) has no effect.

For subscriber processes to reference publisher processes's message objects through this method, all message objects must reside on the heap.
This requirement is naturally satisfied for ROS~2 messages because they consist only of primitive types and arrays (equivalent to standard arrays or \lstinline|std::vector| in C++), without direct pointer handling \cite{ros_msg}.
Conforming to the memory allocation API shown in \cref{fig-api} ensures all ROS~2 message objects are allocated on the heap.
While polymorphic subtype objects typically contain a hidden vtable pointer in process-local memory—requiring techniques like DVT/GVT \cite{el2017savi} and IVT \cite{wang2019ivt} for inter-process transfer—ROS~2 message objects are non-polymorphic, eliminating this concern.

\vspace{-1mm}
\subsection{Shared Metadata Transactionality} \label{subsection-kernel-module}
\begin{figure}[h]
  \centering
  \includegraphics[width=1.00\linewidth]{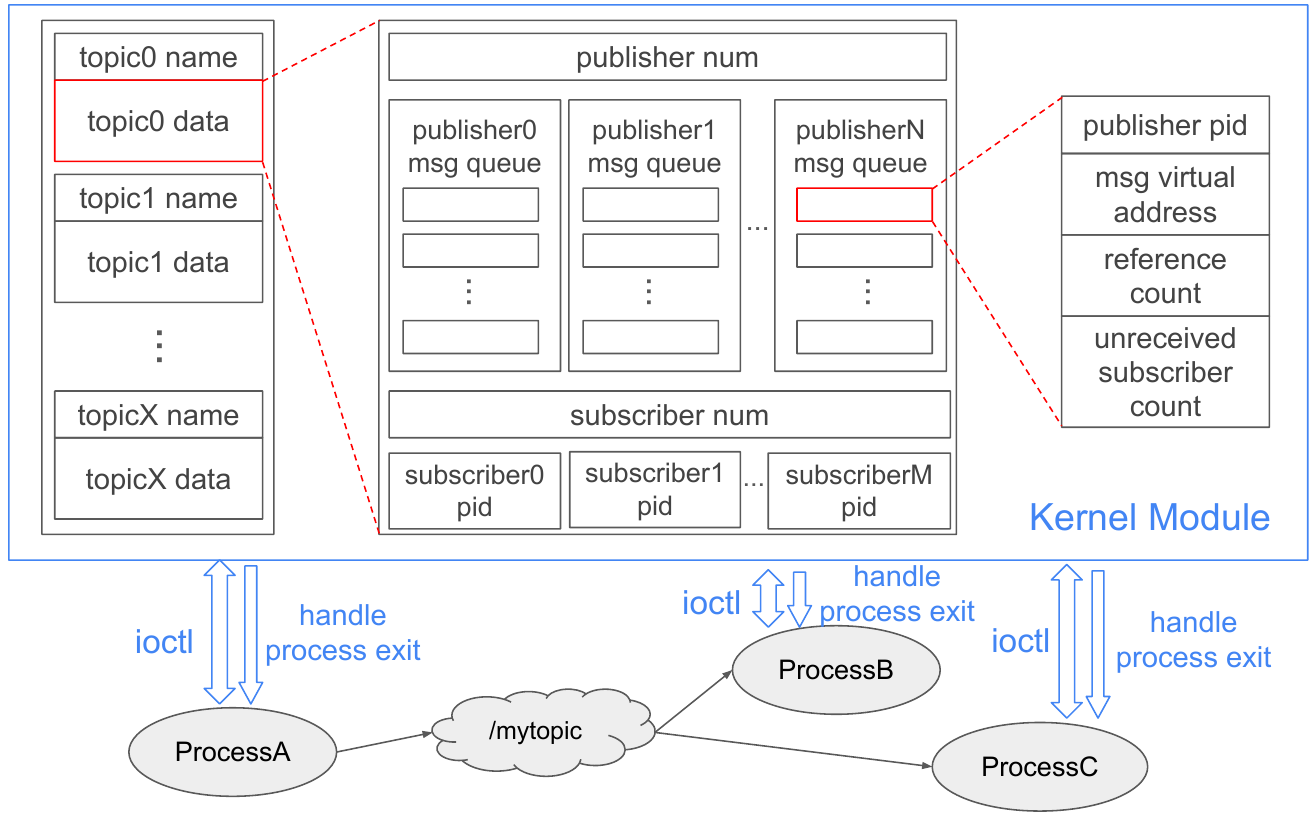}
  \caption{\TOOLNAME{} kernel module}
  \label{fig-kmod}
  \vspace{-2mm}
\end{figure}

\vspace{-1mm}
The management of publish/subscribe communication metadata, such as message virtual addresses and reference counts, is implemented in a kernel module (\cref{fig-kmod}).
This module maintains topic-specific metadata and message queues for each publisher.
Queue entries contain virtual addresses for inter-process message transfer, reference counters, and subscriber receipt tracking information, which are essential for determining message object release timing.
The processes update these shared data through \textit{ioctl} syscall, while the kernel module hooks process exits and maintains data consistency even when a process unexpectedly terminates.

\vspace{-1mm}

The key rationale for kernel-based metadata management is to ensure the transactional integrity of shared data operations. To prevent data corruption, metadata operations must either complete atomically or support proper rollback mechanisms. For implementation simplicity, \TOOLNAME{} adopts the former approach by implementing metadata operations as kernel module system calls, ensuring that updates complete regardless of process termination timing.
Alternatively, the mechanism could be implemented as a user-space daemon instead of a kernel module, which would require rollback functionality to prevent data corruption.

\vspace{-1mm}

\subsection{Timing to Deallocate Messages}
\begin{figure}[h]
  \centering
  \includegraphics[width=1.00\linewidth]{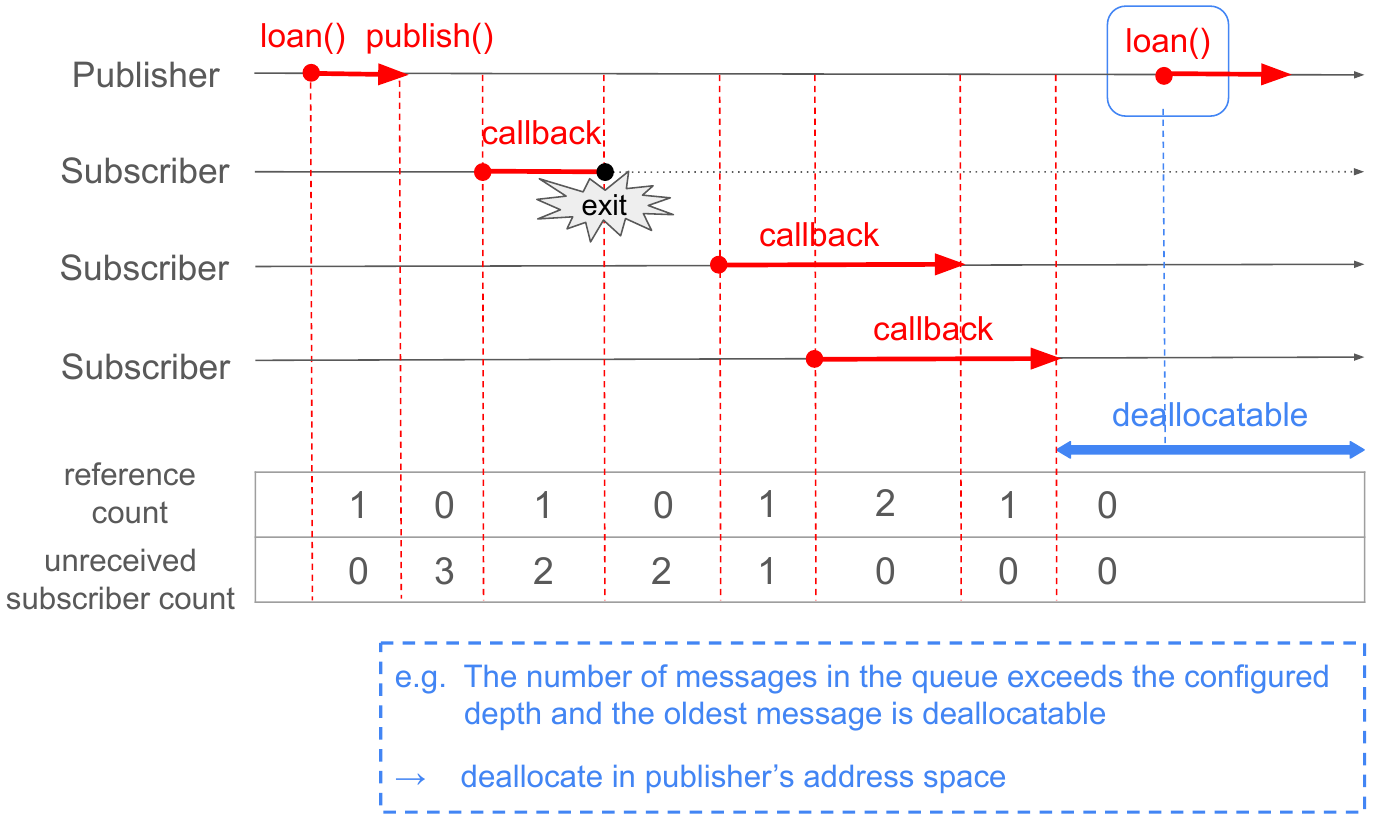}
  \caption{The timing of deallocating \TOOLNAME{} message data}
  \label{fig-deallocate-timing}
\end{figure}

\vspace{-1mm}
As shown in \cref{fig-deallocate-timing}, an allocated message object can be freed when both its \textit{reference count} and \textit{unreceived subscriber count} drop to zero within the allocating process.
While conventional heap memory can be deallocated when its reference count reaches zero (e.g., with \lstinline|std::shared_ptr|), message objects in \TOOLNAME{} require a different approach due to their multi-process accessibility.
\TOOLNAME{} implements its own smart pointer (\cref{fig-api}) that manages both a \textit{reference count} and an \textit{unreceived subscriber count} within the kernel module (\cref{fig-kmod}).
The smart pointer increments the reference count upon creation or duplication and decrements it upon destruction.
Memory deallocation occurs when both counts reach zero, and can only be executed by the publisher process that initially allocated the message, according to its release timing policy.
As described in \cref{subsection-kernel-module}, the kernel module monitors process termination and updates both the \textit{reference count} and \textit{unreceived subscriber count} accordingly to maintain their validity even when processes terminate unexpectedly.

\subsection{Cooperate with Existing Middleware} \label{subsection-bridge}
\begin{figure}[bt]
  \centering
  \includegraphics[width=1.00\linewidth]{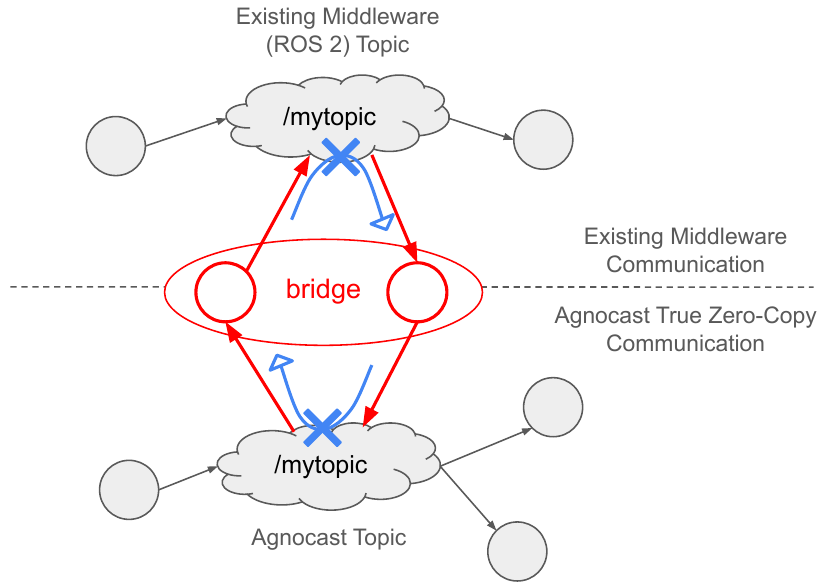}
  \caption{Relaying between \TOOLNAME{} communication and existing middleware communication}
  \label{fig-broker}
  \vspace{-3mm}
\end{figure}

\vspace{-1mm}
To meet Requirement~3 in \cref{subsection-requirements}, \TOOLNAME{} launches a \textit{bridge} process per topic, as described in \cref{fig-broker}, to mediate between existing middleware (e.g. ROS~2) communications and \TOOLNAME{} communications.
This \textit{bridge} process publishes data received from an existing middleware topic to an \TOOLNAME{} topic while it publishes data obtained from an \TOOLNAME{} topic to an existing middleware topic.
To prevent infinite loops, the \textit{bridge's} subscription callback ignores messages originating from itself in both communication paths.
Through this mechanism, messages published in either communication space reach the corresponding subscription callbacks in the other communication space.
This satisfies Requirement~3 in \cref{subsection-requirements}, enabling true zero-copy communication for the targeted communication edges while maintaining the functionality of the whole system.

\section{EVALUATION}

In this chapter, we evaluate \TOOLNAME{}'s performance for ROS~2 under the environment specified in \cref{table-specification}.
Some experimental measurements are represented using violin plots, with vertical bars indicating the maximum and minimum values in the datasets.

\begin{table}[h]
  \centering
  \caption{Evaluation platform.}
  \vspace{0mm}
  \label{table-setup}
  \Large 
  \resizebox{0.48\textwidth}{!}{
    \begin{tabular}{p{5cm} p{10cm}}
      \toprule
      Processor & Intel (R) Xeon (R) E-2278GE CPU @ 3.30 GHz\\
      Number of Cores & 8 * 2 (Hyper-Threading Enabled)\\
      Memory & 32 GB\\
      OS Kernel & Linux Kernel 6.2\\
      OS Distribution& Ubuntu 22.04 LTS\\
      ROS~2 & Humble \\ 
      DDS & Cyclone DDS \\ 
      \bottomrule
    \end{tabular}
  }
  \label{table-specification}
  \vspace{-3mm}
\end{table}

\subsection{IPC Efficiency} \label{subsection-ipc-efficiency}
To evaluate the efficiency of \TOOLNAME{} IPC mechanisms against other ROS~2 IPC mechanisms, we conduct communication latency measurements between publisher and subscriber nodes running in separate processes.
We compare standard ROS~2 communication via CycloneDDS, the existing practical true zero-copy middleware IceOryx, and our proposed \TOOLNAME{} approach.
The evaluation setup consists of one-to-one publisher-subscriber pairs, with threads scheduled by the Completely Fair Scheduler (CFS) under no additional system load, measuring communication latency across varying message sizes (1KB, 10KB, 100KB, and 1MB).
For each configuration, we plot the latency distribution of 1,000 message transmissions performed consecutively at 100ms intervals, excluding the first 10 measurements to avoid initialization effects.
To demonstrate practical applicability in real-world ROS~2 applications, we focus on the \textit{PointCloud2} message type communication, which represents a significant bottleneck in Autoware as discussed in \cref{section-motivation}.
Since IceOryx only supports true zero-copy communication for static-sized message types, we evaluate it under two conditions: using the unsized \textit{PointCloud2} message type and using an equivalent static-sized custom message type.
Notably, we exclude existing true zero-copy approaches such as TZC and LOT from our comparison as they fail to meet the requirements outlined in \cref{subsection-requirements}: they do not support unsized message types and require significant modifications to existing ROS~2 application code and the use of message type implementations specialized for their respective methods, making them impractical for real-world adoption.

\vspace{-3mm}

\begin{figure}[h]
  \centering
  \includegraphics[width=1.00\linewidth]{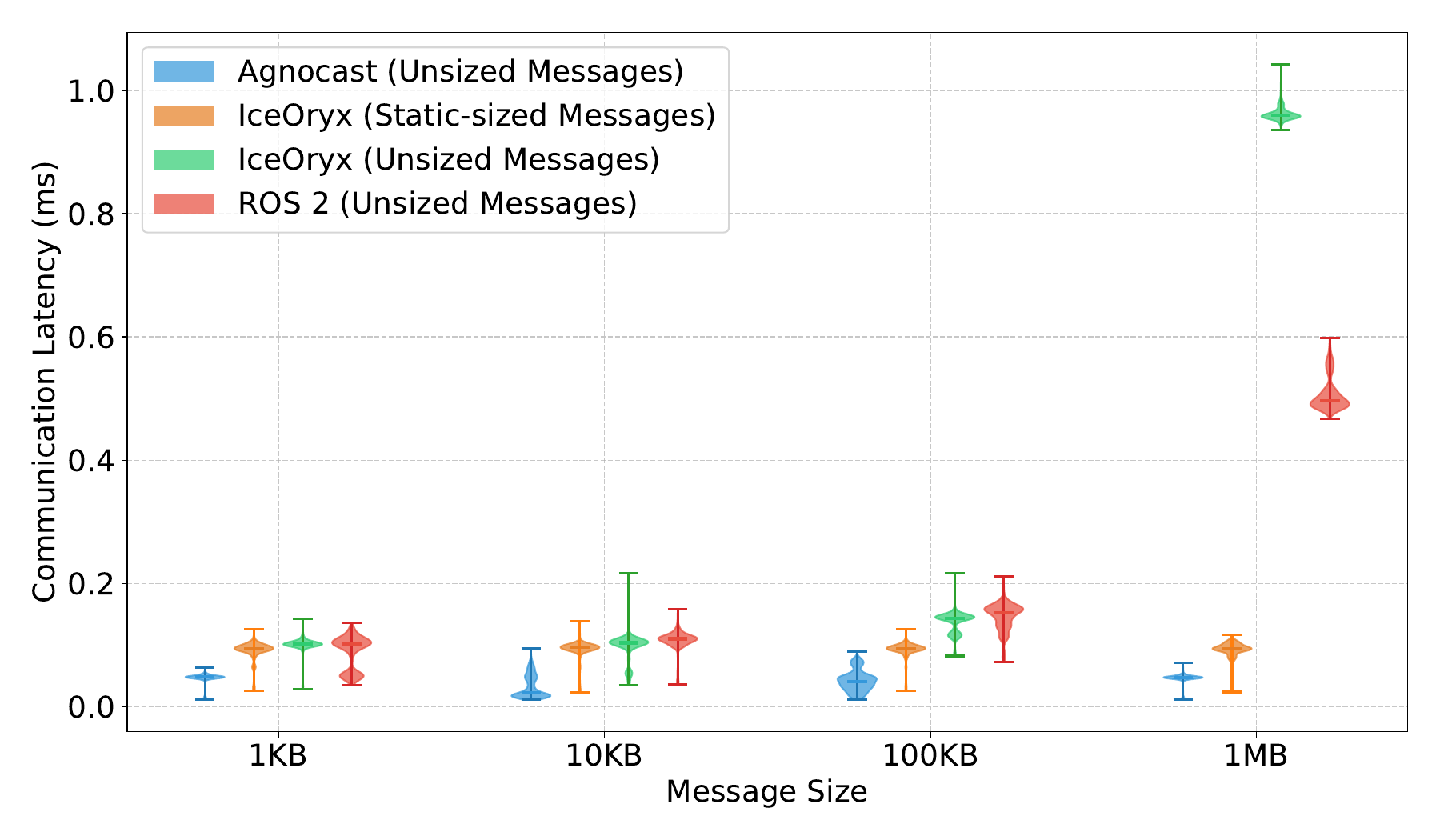}
  \caption{Comparison of communication latency (\TOOLNAME{}, IceOryx and ROS~2)}
  \label{fig-latency-with_sizes}
\end{figure}

\cref{fig-latency-with_sizes} plots the communication latency from message publication to subscriber callback invocation across different message sizes for \TOOLNAME{} and the compared IPC mechanisms. 
\TOOLNAME{} maintains consistent communication latency regardless of message size, even for unsized message types like \textit{PointCloud2}.
For IceOryx, when using static-sized message types, true zero-copy communication effectively maintains constant latency across message sizes.
However, when communicating unsized message types, IceOryx incurs transparent serialization costs to shared memory and deserialization costs from shared memory, leading to increased latency as message size grows.
While IceOryx handling unsized message types was expected to show lower IPC costs than CycloneDDS-based communication by avoiding copy costs in the network stack, measurements revealed higher latency for IceOryx; we do not investigate the detailed cause here.
The results demonstrate that \TOOLNAME{} achieves true zero-copy communication even for unsized message types, showing a particular advantage over IceOryx—the only realistic ``competitor’’—at larger message sizes.

\subsection{IPC Performance Stability under CPU load}
Next, we evaluate the stability of IPC performance under CPU load by measuring the communication latency of \TOOLNAME{} and the compared IPC mechanisms while applying system load using \textit{stress-ng}~\cite{stress-ng}.
We conduct experiments under the same conditions as \cref{subsection-ipc-efficiency}, except that the message size is fixed at 100KB.
We vary the CPU load utilization from 0\% to 90\% (0\%, 30\%, 60\%, and 90\%) by creating \textit{stress-ng} processes equal to the number of logical processors, with each process's CPU utilization set to the target load condition.
In this experiment, we set the subscriber's callback thread scheduling policy to \textit{SCHED\_FIFO} to eliminate the impact of the subscriber's runqueue latency on the communication latency measurements.
When we define communication latency as the time from message publication to subscriber callback invocation, it can be broken down into two elements: the time from message publication until the subscriber's callback thread becomes ready, plus the time from the callback thread becoming ready until it is scheduled and begins execution.
While the latter element had minimal impact in \cref{subsection-ipc-efficiency}'s experiments due to the absence of CPU load, it becomes significant in the current experiments with CPU load.
From the perspective of real-time scheduling problem based on the Directed Acyclic Graph (DAG) task model, we are only interested in the former element (the delay from a DAG's preceding subtask completion until the subsequent subtask becomes ready for execution), hence we adopt this workaround to effectively eliminate the latter element's influence.

\vspace{-3mm}

\begin{figure}[h]
  \centering
  \includegraphics[width=1.00\linewidth]{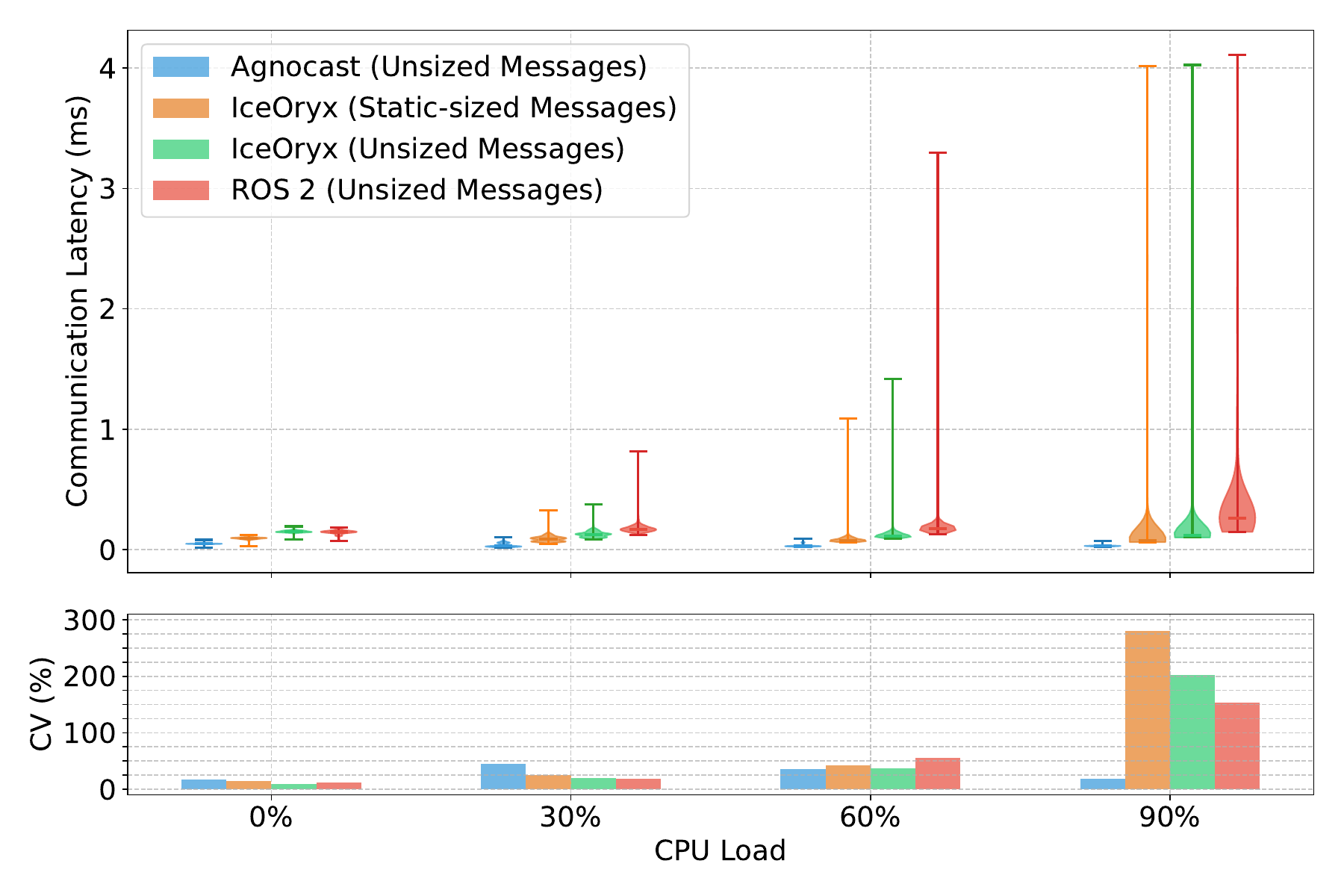}
  \caption{Stability and Real-time Performance under CPU load (Message size is fixed at 100KB)}
  \label{fig-robustness-real-time}
\end{figure}

\cref{fig-robustness-real-time} shows the communication latency measurements for \TOOLNAME{} and the compared IPC mechanisms under different CPU utilization levels.
Here, the communication latency is measured as the time from message publication until the subscriber's callback becomes ready for execution.
Additionally, to evaluate real-time characteristics, we present the coefficient of variation (CV) for each experimental condition.
\TOOLNAME{} maintains high real-time performance even under 90\% CPU utilization.
At 30\% CPU utilization, we observe that CPU usage is not always evenly distributed across cores, likely creating conditions conducive to frequent core migrations, which explain the higher CV compared to other conditions.
For the other IPC mechanisms, the CV increases with CPU utilization, indicating degradation in real-time performance under higher CPU load.
For IceOryx, this is attributed to scheduling delays of the daemon threads that manage IPC operations.
For ROS~2 communication via CycloneDDS, the impact comes from scheduling delays of various involved threads, including DDS threads.
The results demonstrate that \TOOLNAME{} maintains stable IPC performance and exhibits high real-time capabilities even under high CPU utilization.

\subsection{Bridge overhead}
We evaluate the overhead introduced by the ``bridge'' process described in \cref{subsection-bridge}, which mediates communication between \TOOLNAME{} and ROS~2.
For any given topic, the ``bridge'' mechanism is unnecessary when all participants (publishers and subscribers) use either \TOOLNAME{} or ROS~2 exclusively.
However, the ``bridge'' becomes essential when the topic includes a mix of \TOOLNAME{} and ROS~2 participants that need to communicate.
As shown in \cref{fig-broker}, there are two bridging routes: ``\TOOLNAME{} publisher → bridge → ROS~2 subscriber'' and ``ROS~2 publisher → bridge → \TOOLNAME{} subscriber'', and we compare the bridging overhead of each route against standard ROS~2 communication.
The experimental conditions are identical to those in \cref{subsection-ipc-efficiency}, except for the different comparison targets.

\vspace{-3mm}

\begin{figure}[h]
  \centering
  \includegraphics[width=1.00\linewidth]{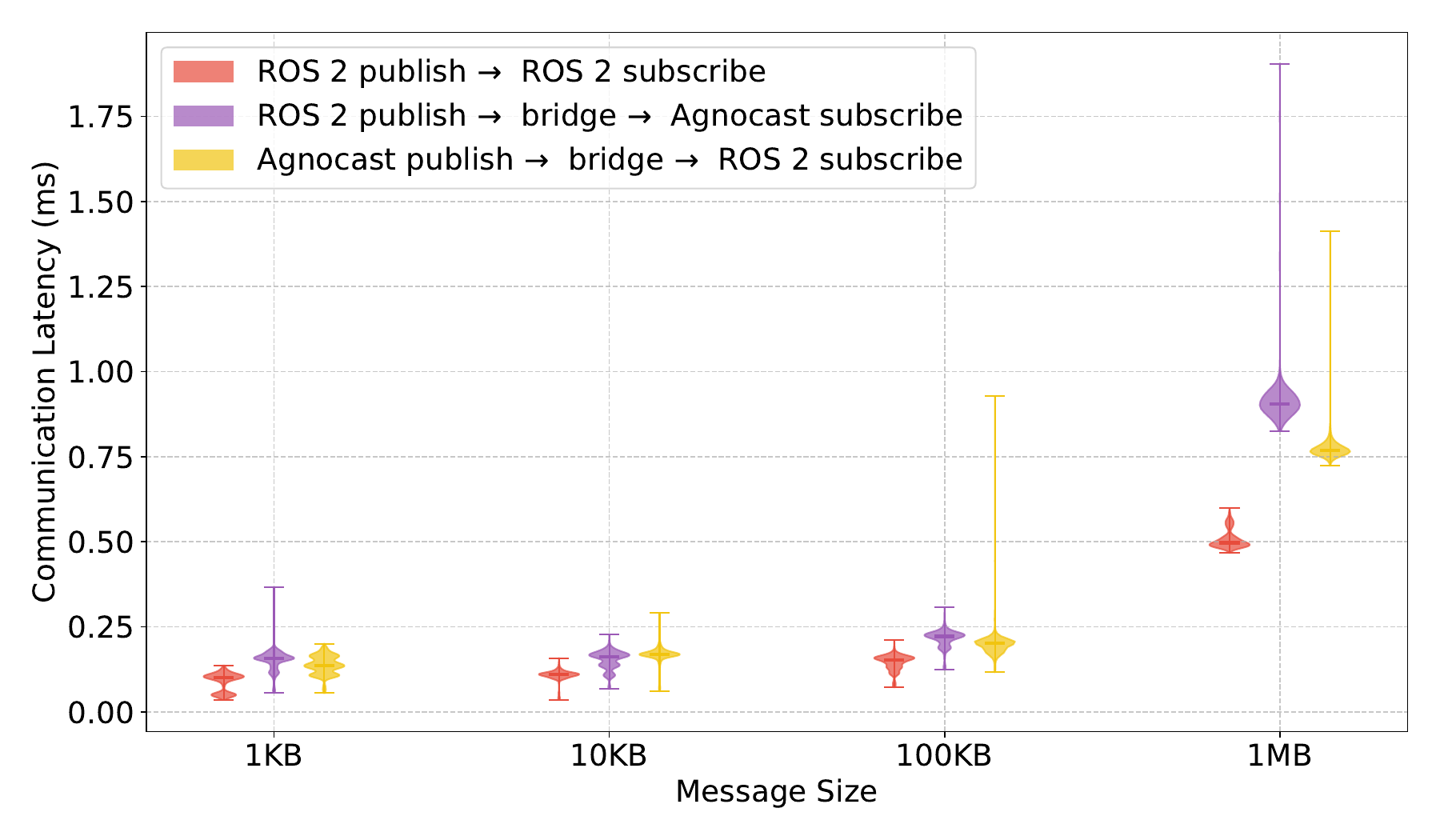}
  \caption{Bridge overhead}
  \label{fig-eval-bridge}
  \vspace{-2mm}
\end{figure}

\cref{fig-eval-bridge} plots the time from message publication to subscriber callback invocation across different message sizes for standard ROS~2 communication and the compared bridge-mediated IPC routes.
Compared to standard ROS~2 communication, bridge-mediated communication incurs communication overhead proportional to message size, with overhead ranging from 0.25 to 0.50 ms for 1MB messages.
This overhead arises from additional message serialization costs in the ``\TOOLNAME{} publisher → bridge → ROS~2 subscriber'' route and additional message copy costs in the ``ROS~2 publisher → bridge → \TOOLNAME{} subscriber'' route.
When introducing the bridge mechanism for incremental migration of publishers/subscribers to \TOOLNAME{} or for supporting inter-host communication, these overhead costs need to be considered for acceptability.

\subsection{Demonstrate on Autoware PointCloud Preprocessor}

\begin{figure}[h]
  \centering
  \includegraphics[width=1.00\linewidth]{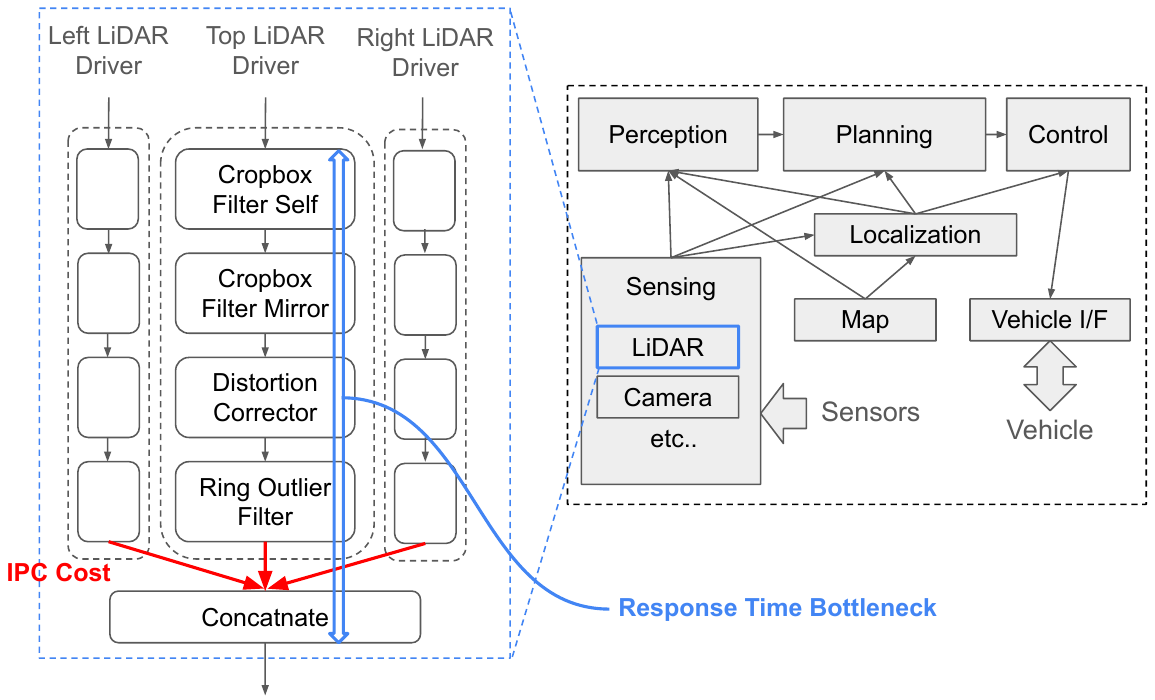}
  \caption{Overview of Autoware's functional units and LiDAR preprocessing, showing the message flow between components and the sequential preprocessing nodes for LiDAR data.}
  \label{fig-autoware}
\end{figure}

To evaluate response time improvements in actual autonomous driving systems, we apply \TOOLNAME{} to one of the performance bottlenecks in Autoware from a response time perspective.
As shown in \cref{fig-autoware}, Autoware forms a comprehensive autonomous driving system through message exchanges between multiple functional units (Sensing, Map, Localization, Perception, Planning, Control, and Vehicle Interface).
Each functional unit comprises multiple ROS~2 nodes, which communicate with other nodes through publish/subscribe messaging.
Within the Sensing functionality, which preprocesses various sensor data, the LiDAR preprocessing unit consists of four preprocessing nodes connected in series for each LiDAR.
The pointcloud data from all LiDARs is finally merged and published to the entire system.
In this evaluation, we focus on measuring the response time of this LiDAR preprocessing functionality.
According to \cite{li2024data}, timing constraints in autonomous systems consist of three components: Freshness (latency from the occurrence of a physical phenomenon to its processing), Consistency (coherence between the programmatic observation order and physical observation order), and Stability (jitter in periodic tasks).
This study contributes to enhancing the system's capability to meet the Freshness constraint.

In this evaluation, we compare the response times before and after applying \TOOLNAME{} to the LiDAR Preprocessing functionality shown in \cref{fig-autoware} in Autoware version 2024.07.
We use the rosbag (a data format designed for recording and replaying sensor data) publicly available in \cite{autoware_rosbag}.
The measurement of the response time is conducted using CARET \cite{kuboichi2022caret}.
This rosbag contains data from three LiDARs (Top, Left, and Right), with the Top LiDAR processing being the response time bottleneck and thus our measurement target.
While the four preprocessing nodes (\textit{cropbox filter self}, \textit{cropbox filter mirror}, \textit{distortion corrector}, \textit{ring outlier filter}) for each LiDAR are placed in the same process by \textit{ComponentContainer}, eliminating communication costs between them, the \textit{concatenate} node resides in a separate process, incurring IPC costs.
Therefore, we apply \TOOLNAME{} to this topic (between \textit{ring outlier filter} and \textit{concatenate}).
The Top LiDAR preprocessing becomes a bottleneck because its pointcloud data size is significantly larger than those of the Left and Right LiDARs (Top LiDAR data is in the MB order, while the other two are in the KB order).
The reason for not placing all LiDAR preprocessing nodes, including the concatenate node, in the same process is to maintain fault isolation.

\begin{figure}[h]
  \centering
  \begin{subfigure}[b]{1.00\linewidth}
    \centering
    \includegraphics[width=1.00\linewidth]{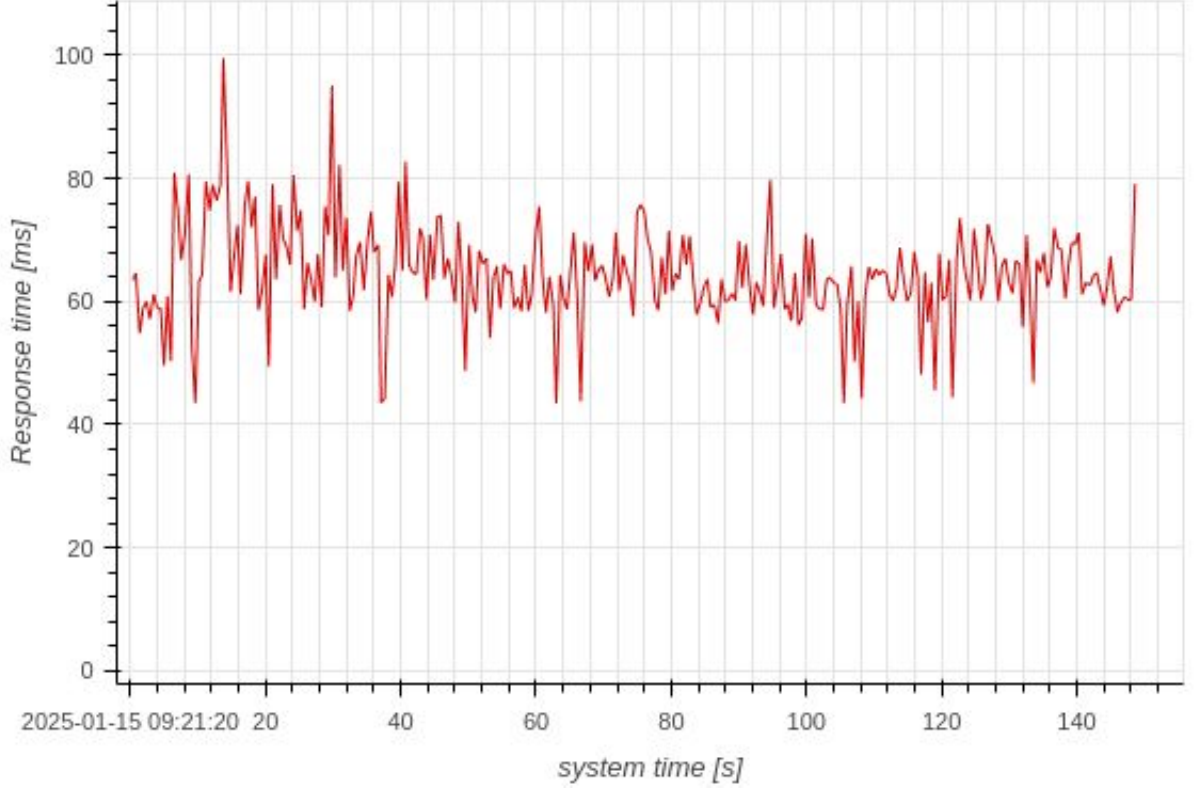}
    \label{fig-eval-autoware-before}
  \end{subfigure}
  
  \begin{subfigure}[b]{1.00\linewidth}
    \centering
    \includegraphics[width=1.00\linewidth]{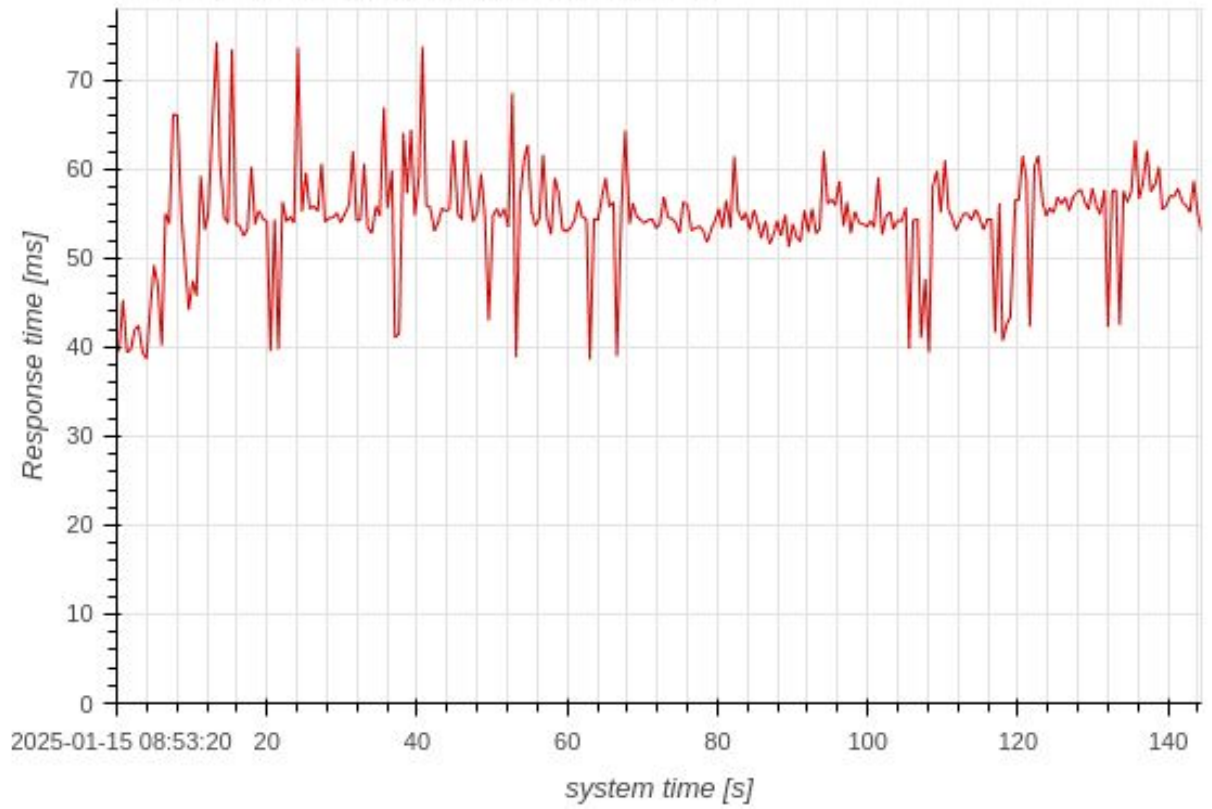}
    \label{fig-autoware-b}
  \end{subfigure}
  \caption{Response time comparison in Top LiDAR preprocessing in Autoware: without (upper) and with (lower) \TOOLNAME{}.}
  \label{fig-eval-autoware}
\end{figure}

\cref{fig-eval-autoware} plots the response time series between across Top LiDAR \textit{cropbox filter self} node to the \textit{concatenate} node (depicted as ``Response Time Bottleneck'' in \cref{fig-autoware}), comparing the performance before and after applying \TOOLNAME{}.
As shown in the figure, applying \TOOLNAME{} to just a single topic results in significant response time improvements.
Before applying \TOOLNAME{}, the average response time is 64.5ms with a maximum value of 99.5ms.
After applying \TOOLNAME{}, these values improve to an average of 54.5ms and a maximum of 74.3ms, achieving approximately 16\% improvement in average response time and 25\% improvement in maximum response time.
Through \TOOLNAME{}, we successfully apply true zero-copy IPC for the unsized message type \textit{PointCloud2}, with its significant impact demonstrated in this real-world application.

\section{RELATED WORK}
While several studies have explored true zero-copy communication through shared memory for unsized message types in ROS~1 and ROS~2, none has been practical for real-world ROS~2 projects.
TZC (Towards Zero Copy) \cite{wang2019tzc} introduces a method of partial serialization.
This approach communicates array data with dynamic size in zero-copy on shared memory, while transmitting the remaining fields through socket communication.
Implementing TZC's approach necessitates the creation of original classes with TZC-specific implementations for each message type, preventing the use of default ROS~2 messages.
This approach also obliges us to write user code with library constraints in mind, such as pre-determining array sizes before calling memory allocation functions.
While TZC supports zero-copy IPC for messages with dynamic size at runtime, this only means that array sizes can be determined dynamically at the beginning of callback functions in the source code.
For example, it does not provide full freedom to call arbitrary functions on \lstinline|std::vector| properties of messages.
LOT (Low Overhead Transport) \cite{iordache2021smart} employs the \lstinline|boost::interprocess| shared memory feature to achieve true zero-copy communication in ROS~1, utilizing custom memory allocation on shared memory.
However, this approach is impractical for developing a usable library in ROS~2, as it necessitates converting ROS~2 message types into LOT-specific formats, leading to the same challenges encountered with TZC.
Like TZC, LOT also requires developers to write their code with library-specific constraints in mind.

In ROS~2, middleware enabling true zero-copy communication for static-sized message types is already fully practical.
IceOryx \cite{pohnl2022middleware, pohnl2023shared} can communicate ROS~2 messages with static size in true zero-copy.
IceOryx is practical for deployment, offering direct access through the ROS~2 C++ API, especially when integrated with CycloneDDS \cite{cyclonedds}.
However, it becomes impractical for applications that heavily depend on unsized message types, as it requires both static array size specification in message files before compilation and extensive modifications to code that assumes unsized message types throughout the codebase.
A method also exists for zero-copy communication with peripheral devices, such as GPUs \cite{bell2023hardware}.
However, these methods are similarly constrained to handling messages of only static size.

Transferring polymorphic subtype objects across processes presents challenges due to vtable pointer invalidation in different processes.
DVT/GVT \cite{el2017savi} addresses this by maintaining consistent vtable addresses or using hash tables, while IVT \cite{wang2019ivt} converts vtable pointers to universal indexes with process-specific correspondence tables.
However, this challenge is irrelevant for ROS~2 message structures, which don't contain polymorphic subtype objects.

Although not directly aimed at achieving true zero-copy on shared memory, there are ongoing research efforts dedicated to lowering the costs of serialization and deserialization, which is the primary bottleneck of message-passing IPC~\cite{nishimura2021raplet}.
Flat-Data \cite{flatdataapi}, as adopted by Connext DDS \cite{connext}, and Google's FlatBuffer \cite{flatbuffer} enable direct data construction, bypassing the need for serialization and deserialization.
However, this approach necessitates using specific APIs for message construction.
Conversely, ROS-SF \cite{wang2022ros} introduces a new format, SFM (Serialization-free Format), which addresses the issue by ensuring the memory layout of serialization-free messages is identical to that of standard messages.
This approach eliminates the need for code changes but requires modifications to the tool that automatically generates code for message structures from the IDL.
ATSA (Adaptive Two-layer Serialization Algorithm) \cite{jiang2020message} simplifies operations and reduces costs by shifting serialization tasks from the \textit{rmw} layer to the programming language layer.

\section{Discussion} \label{section-discussion}
\subsection{Drawbacks of mapping heap to shared memory.}
While subscriber processes can access the publisher process's heap, it is protected as read-only by MMU, preventing any corruption of the publisher process's heap space.
\TOOLNAME{} should not be used for topics where the subscriber could be potentially malicious, such as when the subscriber is a third-party node with unclear implementation details.
The risk of message data being modified while being accessed by the subscriber can be eliminated without using locking mechanisms.
This is because by accepting only rvalue references as arguments to the \textit{publish()} function, there is no way for the message content to be modified after \textit{publish()} is called.

Mapping the heap onto shared memory does not worsen the total size of allocations across the system.
However, it can lead to deterioration in reclaiming physical pages that correspond to virtual memory ranges freed in the heap layer.
In other words, even if a shared physical page is no longer used as heap, the Linux kernel cannot reclaim it.
This arises from the fact that: 1) in the case of private anonymous pages, when the page table entry (PTE) referencing that physical page is gone, the kernel automatically reclaims that physical page; 2) in the case of shared pages, even if all processes referencing the physical page discard their PTEs and the reference count to the physical page becomes zero, the physical page is not reclaimed.
Issuing \textit{MADV\_REMOVE} via the \textit{madvice} syscall will allow the kernel to reclaim it for reuse.
However, it is difficult to determine the reclaimable memory range in the middleware layer, and it must be issued from the application code that understands the meaning of the memory range (e.g., memory for IPC messages or other uses).

To mitigate or resolve the aforementioned drawbacks, minimizing the use of shared memory is crucial.
Since \TOOLNAME{} aims to transfer messages with true zero-copy in publish/subscribe communication, there is inherently no need to place other memory regions on shared memory.
In the prototype \TOOLNAME{}, all heap memory is provisionally placed on shared memory due to the difficulty in determining the application-level semantics only by simply observing the \textit{malloc} and \textit{free} functions intercepted via \textit{LD\_PRELOAD}.
For future work in the production version, we will develop a feature to minimize the use of shared memory to only the necessary parts.

\subsection{Limitations on message types handled by \TOOLNAME{}}
While \TOOLNAME{} is an IPC technology capable of true zero-copy transfer of arbitrary ROS~2 message types, it cannot handle arbitrary objects.
Specifically, it cannot handle message types that have a user-assignable pointer type as a property.
This is because for true zero-copy communication, all message objects must be placed in shared memory.
If a message property has an arbitrarily modifiable pointer, application programmers can potentially set an address that points to memory other than shared memory.
Since ROS~2 message types do not have properties of user-assignable pointers, they fall within the scope of what \TOOLNAME{} can handle.

\section{CONCLUSION AND FUTURE WORK}
This paper introduces \TOOLNAME{}, a new IPC technology that enables true zero-copy IPC with minimum constraints on application code, even when unable to specify memory allocation strategies in application code.
\TOOLNAME{} consists of two key concepts: an idea of directly mapping the heap to shared memory, and a novel smart pointer technology that efficiently manages object lifetimes when objects are accessed across multiple processes.
\TOOLNAME{} can handle any ROS~2 message, including unsized message types, while minimizing code modifications in existing projects.
Our evaluation demonstrates both the superior IPC performance of \TOOLNAME{} compared to existing technologies and its practical utility in a practical autonomous driving system.
As the first open-source true zero-copy IPC middleware to fulfill the deployment needs of large-scale, real-world ROS~2 systems, \TOOLNAME{} represents a significant advancement in ROS~2 communication technologies.
Furthermore, its novel approach to IPC extends beyond ROS~2 specific improvements, contributing to the advancement of general IPC technologies.
A prototype version of \TOOLNAME{} is available as open-source\footnote{\url{ https://github.com/sykwer/agnocast.}}.

As future work, we are developing a production version of \TOOLNAME{} for real-world ROS~2 projects such as Autoware, supporting various standard ROS~2 features such as ROS~2 clients and servers.
We also plan advanced features, including dynamically selecting a virtual address range for shared memory when the intended range is already occupied.

\section{Acknowledgment}
This research is based on results obtained from a project, JPNP21027, subsidized by the New Energy and Industrial Technology Development Organization (NEDO).
Green Innovation Fund Projects / Development of In-vehicle Computing and Simulation Technology for Energy Saving in Electric Vehicles.

\addtolength{\textheight}{-12cm}   

\bibliographystyle{IEEEtran} 
\bibliography{references}     

\end{document}